\documentstyle[A4,epsf,12pt]{article}

\begin{document}

\overfullrule=0pt
\parskip=3pt
\font\bigrm=cmr10 at 14truept
\overfullrule 0pt

\catcode`\@=11
\def\undefine#1{\let#1\undefined}
\def\newsymbol#1#2#3#4#5{\let\next@\relax
 \ifnum#2=\@ne\let\next@\msafam@\else
 \ifnum#2=\tw@\let\next@\msbfam@\fi\fi
 \mathchardef#1="#3\next@#4#5}
\def\mathhexbox@#1#2#3{\relax
 \ifmmode\mathpalette{}{\m@th\mathchar"#1#2#3}%
 \else\leavevmode\hbox{$\m@th\mathchar"#1#2#3$}\fi}
\def\hexnumber@#1{\ifcase#1 0\or 1\or 2\or 3\or 4\or 5\or 6\or 7\or 8\or
 9\or A\or B\or C\or D\or E\or F\fi}

\font\tenmsa=msam10
\font\sevenmsa=msam7
\font\fivemsa=msam5
\newfam\msafam
\textfont\msafam=\tenmsa
\scriptfont\msafam=\sevenmsa
\scriptscriptfont\msafam=\fivemsa
\edef\msafam@{\hexnumber@\msafam}
\mathchardef\dabar@"0\msafam@39

\newsymbol\leqslant 1336
\newsymbol\geqslant 133E

\renewcommand{\leq}{\leqslant}
\renewcommand{\geq}{\geqslant}

\font\tenmsb=msbm10
\font\sevenmsb=msbm7
\font\fivemsb=msbm5
\newfam\msbfam
\textfont\msbfam=\tenmsb
\scriptfont\msbfam=\sevenmsb
\scriptscriptfont\msbfam=\fivemsb

\def\msb{\tenmsb\fam\msbfam}
\def\Bbb{\ifmmode\let\next\Bbb@\else
 \def\next{\errmessage{Use \string\Bbb\space only in math mode}}\fi\next}
\def\Bbb@#1{{\Bbb@@{#1}}}
\def\Bbb@@#1{\fam\msbfam#1}

\def\R{{\Bbb R}}
\def\N{{\Bbb N}}
\def\Z{{\Bbb Z}}
\def\Q{{\Bbb Q}}

\def\1{\hbox{ 1\hskip -3.5pt I}}

\newtheorem{Pro}{Proposition}[section]
\newtheorem{Def}[Pro]{Definition}
\newtheorem{Thm}[Pro]{Theorem}
\newtheorem{Lem}[Pro]{Lemma}
\newtheorem{Rem}[Pro]{Remark}
\newtheorem{Cor}[Pro]{Corollary}

\font\twelve=cmbx10 at 15pt
\font\ten=cmbx10 at 12pt
\font\eight=cmr8
 
\begin{titlepage}

\renewcommand{\thefootnote}{\fnsymbol{footnote}}

\begin{center}



{\twelve EXTENDED SYMBOLIC DYNAMICS IN BISTABLE CML: EXISTENCE AND STABILITY OF
FRONTS}

\vspace{0.3 cm}
\setcounter{footnote}{0}
\renewcommand{\thefootnote}{\arabic{footnote}}

{\bf Ricardo COUTINHO\footnote{Departamento de Matem\'atica, Instituto Superior
T\'ecnico, Av.\ Rovisco Pais 1096, Lisboa Codex} and
Bastien FERNANDEZ\footnote{CNRS Centre de Physique Th\'eorique and Universit\'e
de la M\'editerran\'ee, Luminy F-13288 Marseille Cedex 9}}


\vspace{1.5 cm}

{\bf Abstract}

\end{center}
We consider a diffusive Coupled Map Lattice (CML) for which the local map is 
piece-wise affine and has two stable fixed points. By introducing a 
spatio-temporal coding, we prove the one-to-one correspondence between the set
of global orbits and the set of admissible codes. This relationship is applied 
to the study of the (uniform) fronts' dynamics. It is shown that, for any 
given velocity in $[-1,1]$, there is a parameter set for which the fronts with 
that velocity exist and their shape is unique. The dependence of the velocity 
of the fronts on the local map's discontinuity is proved to be a Devil's 
staircase. Moreover, the linear stability of the global orbits which do not 
reach the discontinuity follows directly from our simple map. For the fronts, 
this statement is improved and as a consequence, the velocity of all the 
propagating interfaces is computed for any parameter. The fronts are shown to 
be also nonlinearly stable under some restrictions on the parameters.
Actually, these restrictions follow from the co-existence of uniform fronts 
and non-uniformly travelling fronts for strong coupling. Finally, these results
 are extended to some $C^{\infty}$ local maps.  

\vspace{1 cm}

\bigskip

\noindent Number of figures: 2

\bigskip

\noindent June 1996




\bigskip

\noindent Submitted to Physica D
\end{titlepage}

\section{Introduction}

The occurrence of interfaces is a very common phenomenon in extended 
systems out of equilibrium.
These patterns appear, in systems with many equilibrium states, as
solutions linking two different equilibria, namely two phases \cite{Cross}. 
When the time
evolution of the system shows the motion of an interface resulting in an 
invasion of one phase into the other, the phenomenon under consideration is 
usually called a front. These structures arise in various experimental systems 
such as reaction-diffusion chemical systems \cite{Showalter}, 
alloy solidification \cite{Levine} and crystal growth 
\cite{Elkinani}. 
\bigskip

From a mathematical point of view, the description of the front dynamics 
using
space-time continuous models is now complete, at least for several 
one-dimensional Partial Differential
Equations \cite{Collet90}. Also, in discrete space and continuous time systems,
i.e.\ in Ordinary Differential Equations, the dynamics of these interfaces 
is well understood \cite{Defontaines,Keener}. However, in space-time discrete
models such as the Coupled Map Lattice (CML), the study is not as 
complete\footnote{The problem of travelling waves has
been investigated in other space-time discrete models such as the chain of 
diffusively coupled maps for which the coupling is 
different from that in CML's \cite{Afraimovich}.};
except for the one-way coupled model \cite{Carretero}.
CML's have been proposed as the simplest models of space-time discrete dynamics
with continuous states and they serve now as a paradigm in the framework of
nonlinear extended dynamical systems \cite{Kaneko93}.
\bigskip

The phenomenology of the interface dynamics differs between discrete 
space and continuous space systems. In the former case,
varying the coupling strength induces a
bifurcation\footnote{This bifurcation is of saddle type and is accompanied
by a symmetry breaking \cite{Bastien95}.} 
from a regime of standing interfaces to a regime of
fronts in which the propagation can be interpreted as a mode-locking 
phenomenon \cite{Defontaines,Carretero}. These effects are assigned to the 
discreteness of space and are well-known in solid-state physics 
\cite{Aubry}.
\bigskip

In previous studies of the piece-wise affine bistable CML, the problem of 
steady interfaces was solved and the fronts' structure was thoroughly 
investigated by means of generalized transfer matrices 
\cite{Bastien95,Laurent}.
Nevertheless, this technique did not allowed us to prove the existence of 
fronts of any velocity, nor to understand clearly their dependence
on the parameters, in particular that of the fronts' velocity.
\bigskip

In the same CML, using techniques employed for the 
piece-wise linear mappings
\cite{Bird,Coutinho96a}, we now prove the one-to-one correspondence between 
the set of orbits that are defined for all times, i.e.\ the global orbits, and 
a set of 
spatio-temporal codes (Section 2 and 3). Our model is locally a 
contraction, hence these orbits are shown to be linearly stable when they never
reach the local map's discontinuity (Section 4). Further, using the orbit-code 
correspondence, the existence of fronts is stated and their velocity is 
computed (Section 5). 
In Section 6, the linear stability of fronts with rational velocity 
is given for a large 
class of initial conditions. In the following, we study the dynamics 
of the propagating interfaces. In particular, their velocity is computed 
for all the parameters (Section 7). Using these results, 
the nonlinear stability, i.e.\ the stability of fronts with respect to 
any kink initial condition, is proved in Section 8 using a method 
similar to the Comparison Theorem \cite{Collet90}. This result holds 
for any rational velocity provided the coupling is small enough. 
We justify such a 
restriction by the existence of non-uniform fronts for 
large coupling which co-exist with the fronts. Finally, some 
concluding remarks are made, in particular we emphasize the extension of these
results to certain $C^{\infty}$ local maps.

\section{The CML and the associated coding}

Let $M\in \R$ be fixed 
and ${\cal M}=[-M,M]^{\Z}$ be the phase space of the CML 
under consideration. The CML is the one-parameter family of maps
$$\begin{array}{rl}
F_{\epsilon}:&{\cal M} \longrightarrow {\cal M}      \\
              & x^t \longmapsto x^{t+1}
, \end{array}$$
where $x^t=\left\{x_s^t\right\}_{s\in\Z}$ is the state of the system at 
time $t$.
This model is required to be representative of the reaction-diffusion dynamics.
Therefore the new state at time 
$t+1$ is given by 
\cite{Kaneko93}:
\begin{equation} \label{CML}
x_s^{t+1}\equiv (F_{\epsilon} x^t)_s=(1-\epsilon)f(x_s^t)+{\epsilon \over 2}
 \left( f(x_{s-1}^t)+f(x_{s+1}^t) \right)\quad \forall s \in \Z .
\end{equation}
Here the coupling strength $\epsilon \in (0,1]$ and we
 choose the local 
map $f$ to be bistable and piece-wise affine \cite{Laurent}:
$$f(x)=\left\{ 
\begin{array}{ccc}
ax+(1-a)X_1&if&x<c\\
ax+(1-a)X_2&if&x\geq c,
\end{array} \right.$$
where $a\in (0,1)$ and $-M\leq X_1<c\leq X_2 \leq M$. These conditions ensure 
the existence 
of the two stable fixed points $X_1$ and $X_2$, the only attractors for $f$.
\bigskip

This local map reproduces qualitatively the
autocatalytic reaction in chemical systems \cite{Showalter}, or the
local force applied to the system's phase in the process of solidification 
\cite{Levine}.
\bigskip

For the sake of simplicity we assume $X_1=0$ and $X_2=1$. 
This is always possible by a linear transformation of the variable.
\bigskip 

For a state $x^t$, the sequence 
$\theta^t=\left\{\theta_s^t\right\}_{s\in\Z}\ $ defined by 
$$\theta_s^t=\left\{ 
\begin{array}{ccc}
0&if&x_s^t<c\\
1&if&x_s^t\geq c,
\end{array} \right.$$
is called the {\bf spatio-temporal code} or the code unless it is
ambiguous.

\section{The orbit-code correspondence}

The study of the orbits, in particular those that exist for all the times 
$t\in\Z$, can be achieved
using their code. In this section, we first compute explicitly the positive
orbits\footnote{i.e.\ the orbits for $t\geq 0$} for any initial 
condition. Then we prove the one-to-one correspondence between the
global orbits and their code.
\bigskip

Notice that the local map can be expressed in terms of the code
$$f(x_s^t)=ax_s^t+(1-a)\theta_s^t.$$ 
By introducing this expression into the CML dynamics, one obtains a 
linear non-homogeneous finite-difference equation for $x_s^t$ 
in which the code only appears in the non-homogeneous term.
Using the Green functions' method, this equation may be solved and the 
solution, as a function of the
code $\left\{\theta_s^k\right\}_{s\in\Z ,t_0\leq k\leq t-1}$ and of
the initial condition $\{x_s^{t_0}\}_{s\in\Z }$, is given by
\begin{equation}\label{CONFIG}
x_s^t=\sum_{k=1}^{t-t_0}\sum_{n=-k}^{k}l_{n,k}\theta_{s-n}^{t-k}+
{a\over 1-a}\sum_{n=-(t-t_0)}^{t-t_0}l_{n,t-t_0}x_{s-n}^{t_0}
\quad \forall t>t_0\ {\rm and}\ s\in\Z ,
\end{equation}
where the coefficients $l_{n,k}$ satisfy the recursive relations
$$l_{n,k}=(1-a)\delta_{1,k}\left((1-\epsilon)\delta_{n,0}+{\epsilon\over 2}
(\delta_{n,1}+\delta_{n,-1})\right) \quad \forall n\in\Z\ {\rm and}\ k\leq 
1,\footnote{$\delta_{n,k}$ is the Kroenecker's symbol}$$
and
$$l_{n,k+1}-a(1-\epsilon)l_{n,k}-{a\epsilon\over 2}\left(l_{n+1,k}+l_{n-1,k}
\right)=0 \quad \forall n\in\Z\ {\rm and}\ k\geq1.$$
From the latter, it follows that for $\epsilon\in (0,1)$
$$l_{n,k}=0\quad {\rm iff}\quad |n|>k\ {\rm or}\ n=k=0,$$
and, one can derive the bounds
\begin{equation}\label{BOUND}
0\leq l_{n,k}\leq (1-a)a^{k-1},
\end{equation}
and the normalization condition
$$\sum_{n,k\in\Z }l_{n,k}=1.$$
Further properties of these coefficients are given in Appendix \ref{LNK}.
\bigskip

The study is now restricted to the {\bf global orbits} $\{x^t\}_{t\in\Z}$. 
That is to say, we consider the sequences 
which belong to 
$$\Lambda=\left\{ \{x^t\}_{t\in\Z }\in {\cal M}^{\Z}\ :\ F_{\epsilon}(x^t)=
x^{t+1}\ \forall t\in\Z\right\}.$$
For such orbits, taking the limit 
$t_0\rightarrow -\infty$ and using the bounds (\ref{BOUND}) in (\ref{CONFIG})
leads to the relation:
\begin{equation} \label{ORBIT}
x_s^t=\sum_{k=1}^{+\infty}\sum_{n=-k}^{k}l_{n,k}\theta_{s-n}^{t-k}
\quad \forall s,t\in\Z ,
\end{equation}
which gives the correspondence between these orbits and their code 
as we state now.
\bigskip

By the relation (\ref{ORBIT}), all the orbits in $\Lambda$ stay 
in $[0,1]^{\Z}$. Hence their code can be uniquely computed by 
\begin{equation} \label{CODE}
\theta_s^t=\lfloor x_s^t-c\rfloor +1 \quad \forall s,t \in \Z,
\end{equation}
where $\lfloor .\rfloor$ stands for the floor function\footnote{
i.e.\ $\lfloor x\rfloor\in\Z$ and $x-1<\lfloor x\rfloor\leq x$}.
\bigskip

From (\ref{ORBIT}) and (\ref{CODE}) it follows that for 
any orbit in $\Lambda$, its code must obey the following relation:
$$c-1\leq\sum_{k=1}^{+\infty}\sum_{n=-k}^{k}l_{n,k}\left(
\theta_{s-n}^{t-k}-\theta_s^t\right)<c
\quad \forall s,t\in\Z, $$
which is called the {\bf admissibility condition}. Then, conversely to the 
preceding statement, for a sequence $\theta\in\{0,1\}^{\Z^2}$ that 
satisfies this condition, there is a unique orbit in $\Lambda$, given by 
(\ref{ORBIT}). 
\bigskip

The spatio-temporal coding is an effective tool for the description of all 
the global orbits of the piece-wise affine bistable CML 
\cite{Coutinho96b}. 
These orbits are important since they collect the physical phenomenology
of our model.

\section{The linear stability of the global orbits}
We prove in this section the stability of the orbits in $\Lambda$ 
with respect to small initial perturbations. Let the norm 
$$\| x\|=\sup_{s\in\Z }|x_s |,$$
for $x\in {\cal M}$ and 
$$\left({\rm L}x\right)_s=a\left((1-\epsilon)x_s+{\epsilon\over 2}
(x_{s+1}+x_{s-1})\right) \quad \forall s\in\Z ,$$
be the CML linear component. Notice that ${\rm L}$ is invertible on ${\cal M}$
if $\epsilon<{1\over 2}$ and 
$$\|{\rm L}\|=\sup_{x\neq 0} {\|{\rm L}x\|\over \| x\|}=a.$$
The linear stability is claimed in 
\begin{Pro}\label{STAB1}
Let $\left\{ x^t\right\}_{t\in\N }$ be an orbit such that
$$\exists \delta >0 \ :\ |x_s^t-c|>\delta a^t
\quad \forall t\geq 0,\ s\in\Z .$$
Then for any initial condition $y^0$ in a 
neighborhood of $x^0$, i.e.\ 
$$\|y^0-x^0\| <\delta,$$
we have
$$\|y^t-x^t\| <\delta a^t\quad \forall t\geq 0.$$
\end{Pro}
Equivalently, $x^t$ and $y^t$ have the same code for all times.

\noindent
{\sl Proof:} The relation (\ref{CML}) with the present local map
can be written  
in terms of the operator ${\rm L}$
\begin{equation}\label{LINDY}
x^{t+1}={\rm L}x^t+{1-a\over a}{\rm L}\theta^t
\end{equation}
where $\theta^t=\{\theta_s^t\}_{s\in\Z }$ is the code of $x^t$. Using 
this relation, one shows by induction that the codes of the two orbits 
remains equal for all the times; also using (\ref{LINDY}), the latter implies 
the statement. $\Box$
\bigskip

Notice that this assertion is effective for the orbits in $\Lambda$ that never 
reach $c$, since in this situation $\delta$ can be computed 
(see Proposition \ref{STAB2} below for the case of fronts).
Further, because our system 
is deterministic, when this proposition holds for 
an orbit $\{x^t\}$ and a given initial condition $y$, it cannot hold
for a different orbit $\{\tilde x^t\}$ and the same initial condition $y$, 
unless both these orbits converge to each other.  Hence, using this statement, 
one may be able to determine the (local) basin of attraction for any 
orbit in $\Lambda$ that never reaches $c$.

\section{The existence of fronts}
We now apply the orbit-code correspondence to a particular class of 
travelling wave orbits, namely
the fronts. 
\begin{Def}\label{FRONT}
A {\bf front} with velocity $v$ is an orbit in $\Lambda$ given by
$$x_s^t=\phi(s-vt-\sigma)\quad \forall s,t\in\Z, $$
where $\sigma\in\R$ is fixed and, the {\bf front shape} 
$\phi:\R\longrightarrow\R$, is a right continuous function which obey the 
conditions
\begin{equation}\label{ADMI}
\left\{\begin{array}{ccc}
\phi(x)<c&{\rm if}&x<0\\
\phi(x)\geq c&{\rm if}&x\geq 0.
\end{array}\right.
\end{equation}
\end{Def}
\bigskip

The front shape has the following spatio-temporal behavior:
$$\lim_{x\rightarrow -\infty}\phi(x)=0\quad {\rm and}\quad 
\lim_{x\rightarrow +\infty}\phi(x)=1.$$
In this way, the fronts are actually the travelling interfaces 
as described in the introduction.
Moreover, for any front shape, the front changes by varying $\sigma$; but if
$v={p\over q}$ with $p$ and $q$ co-prime integers, there are only $q$ different
fronts that cannot be deduced from one another by space translations. On the
other hand, when $v$ is irrational, the 
family of such orbits becomes uncountable. (Both these claims are deduced from 
the proof of Theorem \ref{EXIST} below.)
\bigskip

The existence of fronts is stated in
\begin{Thm}\label{EXIST}
Given $(a,\epsilon)$ there is a countable nowhere dense set 
${\rm G}_{a,\epsilon}\subset (0,1]$ such that, for any $c\in (0,1]\setminus
{\rm G}_{a,\epsilon}$, there exists a unique front shape. The 
corresponding front velocity is a continuous function of the parameters with
range $[-1,1]$.
\end{Thm}

In other words, for any velocity $v$, 
there is a parameter set for which the only fronts that 
exist are those of velocity $v$. 
\bigskip

Furthermore, for $c\in {\rm G}_{a,\epsilon}$, no front exists. But, the front 
velocity can be extended to a continuous function for all the values of
the parameters.
\bigskip

Referring to a similar study of a circle's map  
\cite{Coutinho96a}, we point out that, even when no front exists,
numerical simulations show convergence towards a ``ghost'' front. 
Actually, by enlarging the linear stability result, this comment is proved 
(see Proposition \ref{GHOST} below). A
ghost front is a front-like sequence in ${\cal M^{\Z }}$, that is not an 
orbit of the CML and, for which the 
(spatial) shape obeys, instead of (\ref{ADMI}), the conditions 
$$
\left\{\begin{array}{ccc}
\phi(x)\leq c&{\rm if}&x<0\\
\phi(x)> c&{\rm if}&x\geq 0.
\end{array}\right.
$$ 
Ghosts fronts arise in this model because the local map is discontinuous.
Moreover, given the parameters, one can avoid this 
ghost orbit by changing the value of the local map $f$ at $c$.
\bigskip

\noindent
{\sl Proof of Theorem \ref{EXIST}:} 

\noindent
It follows from Definition \ref{FRONT} that if a front with velocity $v$ 
exists, the code associated with this front is given by
$$\theta_s^t=H(s-vt-\sigma)\quad \forall s,t\in\Z ,$$
where $H$ is the Heaviside function\footnote{
$H(x)=\left\{\begin{array}{ccc}
0&{\rm if}&x<0\\
1&{\rm if}&x\geq 0
\end{array}\right.$}.
Hence by (\ref{ORBIT}) the corresponding front shape has the following 
expression:
\begin{equation}\label{SHAPE}
\phi(x)=\sum_{k=1}^{+\infty}\sum_{n=-k}^kl_{n,k}H(x-n+vk)\quad \forall x\in\R.
\end{equation}
Such a function is increasing (strictly if $v$ is irrational), right 
continuous and its discontinuity
points are of the form $x=n-vk\ (k\geq 1,n\in\Z $ and $|n|\leq k)$. Now one
has to prove that, given $(a,\epsilon,c)$, there exists a unique front
velocity such that (\ref{SHAPE}) satisfy the conditions (\ref{ADMI}), i.e.\ 
there exists a unique $v$ such that the code $H(s-vt-\sigma)$ is admissible. 
For this let
$$\eta(a,\epsilon,v)\equiv\phi(0)=\sum_{k=1}^{+\infty}
\sum_{n=-k}^{\lfloor vk\rfloor }l_{n,k},$$
where the continuous dependence on $a$ and $\epsilon$ is included in the 
$l_{n,k}$ which are uniformly summable. It is immediate that 
$\eta(a,\epsilon,v)$ is a strictly increasing ($\epsilon\neq 0$),  
right continuous function of $v$.
Moreover, it is continuous on the irrationals and\footnote{
$g({p\over q}^-)={\displaystyle\lim_{h\rightarrow {p\over q},h<{p\over q}}}
g(h)$}
$$\eta(a,\epsilon,{p\over q})-\eta(a,\epsilon,{p\over q}^-)>0.$$ 
The conditions (\ref{ADMI}) impose 
that $v$ be given by
\begin{equation} \label{VELOCITY}
\bar{v}(a,\epsilon,c)=\min\left\{v\in\R : \eta(a,\epsilon,v)\geq c\right\}.
\end{equation}
Actually, if $\bar{v}(a,\epsilon,c)$ is irrational then 
$$\phi(x)<\phi(0)=c\ {\rm if}\ x<0.$$
If $\bar{v}(a,\epsilon,c)$ is rational we have
$$\phi(x)=\phi(0^-)=\eta(a,\epsilon,{p\over q}^-)\ {\rm if} \ -{1\over q}
\leq x<0,$$
hence either $\phi(0^-)<c$ and the front exists, or $\phi(0^-)=c$ and the
first condition in (\ref{ADMI}) is not satisfied. In the latter case, 
given $(a,\epsilon)$ and $v={p\over q}$, there is a unique $c$ that realizes
$\phi(0^-)=c$. The countability of the values of $c$ for which there is no
front then follows.
\bigskip

In this way, we can conclude that for $a,\epsilon$ and ${p\over q}$ fixed, 
there exists an interval of the parameter $c$ given by 
$$\eta(a,\epsilon,{p\over q}^-)<c\leq \eta(a,\epsilon,{p\over q}),$$
for which the front shape uniquely exists and the front velocity is 
$\bar{v}(a,\epsilon,c)={p\over q}$. Moreover, we have  
$$
{\rm G}_{a,\epsilon}=\left\{\eta(a,\epsilon,{p\over q}^-)\ :\ 
{p\over q}\in (-1,1]\cap \Q\right\}.
$$
\bigskip

The continuity of $\bar{v}(a,\epsilon,c)$ is ensured by the following 
arguments.
Given $\delta>0$, (\ref{VELOCITY}) implies that
$$\eta(a,\epsilon,\bar{v}(a,\epsilon,c)-\delta)<c\ {\rm and}\ 
\eta(a,\epsilon,\bar{v}(a,\epsilon,c)+\delta)>c.$$
Then for $(\tilde a,\tilde\epsilon,\tilde c)$ in a small neighborhood of 
$(a,\epsilon,c)$, one has
$$\eta(\tilde a,\tilde\epsilon,\bar{v}(a,\epsilon,c)-\delta)<\tilde c\ 
{\rm and}\ 
\eta(\tilde a,\tilde\epsilon,\bar{v}(a,\epsilon,c)+\delta)>\tilde c,$$
and again from (\ref{VELOCITY})
$$\bar{v}(a,\epsilon,c)-\delta<\bar{v}(\tilde a,\tilde\epsilon,\tilde c)
\ {\rm and}\ 
\bar{v}(a,\epsilon,c)+\delta\geq\bar{v}(\tilde a,\tilde\epsilon,\tilde c),$$
hence $\bar{v}(a,\epsilon,c)$ is continuous. The range of 
$\bar{v}(a,\epsilon,c)$ follows from the continuity and the values for 
$\epsilon\neq 0$:
$$\eta (a,\epsilon,(-1)^-)=0,\ 
\eta(a,\epsilon,-1)>0\ {\rm and}\ \eta(a,\epsilon,1)=1.$$
 $\Box$
\begin{Rem}
Notice that, from (\ref{VELOCITY}) and the properties of the function $\eta$,
$\bar{v}(a,\epsilon,c)$ is an increasing function of $c$ with range 
$[-1,1]$ when $c$ varies from 0 to 1 ($\epsilon\neq 0$). 
Further, it has the structure of a Devil's staircase.
\end{Rem}
This proof gives also a practical method for computing numerically the
front velocity by inverse plotting $\eta$ versus $v$ (see Figure 1). This 
picture reveals that $\bar{v}(a,1,c)<1$ for some values of $c$. Notice also
that for $c\geq {1\over 2}$, the velocity is non negative. Moreover, the local 
map
has to be sufficiently non-symmetric, i.e.\ $c$ must be sufficiently 
different from ${1\over 2}$, to have travelling fronts. These two 
comments can be stated from the inequality 
$$\eta(a,\epsilon,0)>{1\over 2}.$$

\section{The linear stability of fronts with rational velocity}
In this section, we improve the linear stability result of the 
orbits in $\Lambda$ for the fronts and, also for the ghost fronts.
\bigskip

We now consider the configurations $x\in {\cal M}$ for which the code is 
given by the Heaviside function, namely the {\bf kinks}. Among the kinks, we 
shall use the front's configurations\footnote{From now on, the 
parameters' dependence is removed unless an ambiguity results and, 
we denote by $x$ or by $x^0$ the initial condition of the orbit 
$\{x^t\}_{t\in\N}$.}:
$$\left( R_{\sigma}\right)_s=
\sum_{k=1}^{+\infty} \sum_{n=-k}^k l_{n,k}
H(s-n+\bar{v}k-\sigma)\quad \forall s \in \Z ,$$
where $\bar{v}$ is given by (\ref{VELOCITY}). Further, let the interval 
$${\rm I}_{p\over q}^0=\left(\eta({p\over q}^-),\eta({p\over q})\right),$$
and $\lceil .\ \rceil$ be the ceiling function\footnote{
i.e.\ $\lceil x\rceil\in\Z$ and $x\leq \lceil x\rceil<x+1$}.
The linear stability of fronts is claimed in the following statement.
\begin{Pro}\label{STAB2}
For $c\in {\rm I}_{p\over q}^0$, let $\delta=\min\left\{ c-\eta({p\over q}^-),
\eta({p\over q})-c\right\}$ and $x$ a kink initial condition such that
\begin{equation}\label{CONDIN}
\exists \sigma\in\R : |\left( R_{\sigma}\right)_{s+\lceil\sigma\rceil}
-x_{s+\lceil\sigma\rceil}|\leq\delta a^{-k}\quad\forall k\geq 0,\quad
\forall -2k-1\leq s\leq 2k,
\end{equation}
then $R_{\sigma}^t$ and $x^t$ have the same code for all times and
$$\lim_{t\rightarrow +\infty}\|R_{\sigma}^t-x^t\|=0.$$ 
\end{Pro}
\begin{Rem}\label{T0}
In practice, according to the present phase space, the condition (\ref{CONDIN})
has only to hold for 
$0\leq k\leq t_0$ where $t_0$ is such that $(M+1)a^{t_0}<\delta$.
\end{Rem}

\noindent
{\sl Proof:} Since $x$ is a kink that satisfies (\ref{CONDIN}), we have 
according to the nearest neighbors coupling of the CML:
$$\forall t\geq 0\ 
\left\{\begin{array}{ccl}
x_{s+\lceil \sigma\rceil}^t<c&{\rm if}&s\leq -t-1\\
x_{s+\lceil \sigma\rceil}^t\geq c&{\rm if}&s\geq t.
\end{array}\right.$$ 
Using these inequalities, the condition (\ref{CONDIN}) and the definition of 
$\delta$, one shows by induction, that
$$\forall t\geq 0\ |(R_{\sigma}^t)_{s+\lceil\sigma\rceil}
-x_{s+\lceil\sigma\rceil}^t|\leq\delta a^{-k}\quad\forall k\geq 0,\quad 
\forall -2k-1-t\leq s\leq 2k+t.$$
The latter induces that both the orbits have the same code for all times. 
$\Box $
\bigskip

When it exists, let us define the ghost front of velocity ${p\over q}$ by 
$\{R_{{p\over q}t+\sigma}\}_{t\in\Z }$. For such orbits, the linear 
stability's statement is slightly different than for the fronts:
\begin{Pro}\label{GHOST}
For $c=\eta({p\over q}^-)$, let $\delta=\phi(-{1\over q})-\phi(-{2\over q})$, 
and $x$ a kink initial condition such that
$$
\exists \sigma\in\R : 0<x_{s+\lceil\sigma\rceil}-
\left( R_{\sigma}\right)_{s+\lceil\sigma\rceil}
\leq\delta a^{-k}\quad\forall k\geq 0,\quad\forall -2k-1\leq s\leq 2k,$$
then 
$$\lim_{t\rightarrow +\infty}\|R_{{p\over q}t+\sigma}-x^t\|=0.$$ 
\end{Pro}
The proof is similar to the preceding one noticing that for the ghost 
front
$$R_{{p\over q}(t+1)+\sigma}={1-a\over a}{\rm L}\theta^t+
{\rm L}R_{{p\over q}t+\sigma},$$
where $\theta_s^t=H(s-\lceil {p\over q}t+\sigma\rceil)$ and that $\forall 
t\geq 0$
$$x_{s+\lceil\sigma\rceil}^t>
(R_{{p\over q}t+\sigma})_{s+\lceil\sigma\rceil}\quad \forall s.$$

\section{The interfaces and their velocity}
In this section, we study the
dynamics and the properties of the code of the orbits for which the state is 
a kink for all (positive) times.
\begin{Def}\label{INTERF}
An {\bf interface} is a positive orbit $\{x^t\}_{t \in \N}$
such that
$$\forall t\in \N\ \exists J(x^t)\in\Z\ :\  
\left\{
\begin{array}{ccl}
x_s^t<c&{\rm if}&s\leq J(x^t)-1 \\
x_s^t\geq c&{\rm if}&s\geq J(x^t),
\end{array} \right.$$
where the sequence $\left\{J(x^t)\right\}_{t\in\N}$ is called the {\bf temporal
code}. The velocity of an interface is the limit
$$v=\lim_{t\rightarrow +\infty}{J(x^t)\over t},$$
provided it exists.
\end{Def}

Notice that the front's temporal code is given by
\begin{equation}\label{TEMPCO}
\lceil \bar{v}t+\sigma\rceil\quad \forall t\in \Z,
\end{equation}
namely it is uniform \cite{Laurent}. 
\bigskip

To compare the kinks, the following partial order is considered
\begin{Def}
Let $x,y\in {\cal M}$. We say that $x\prec y$ iff $x_s\leq y_s\quad \forall 
s\in\Z$.
\end{Def}
As a direct consequence of our CML, using this definition, we have
\begin{Pro}\label{ITEOR}
(i) If $x\prec y$ then $F_{\epsilon}x\prec F_{\epsilon}y$.

\noindent
(ii) If $x\prec y$ are two kinks, then $J(x)\geq J(y)$.
\end{Pro}
Moreover, we can ensure a kink to be the initial condition of 
an interface in the following ways
\begin{Pro}\label{CROIS}
Let $x$ be a kink initial condition. 
If $x_s\leq x_{s+1}\ \forall s\in\Z$ or, if 
$\epsilon\leq {2\over 3}$, then $\{x^t\}_{t\in\N }$ is an interface.
\end{Pro}
We will also use the asymptotic behaviour in space of the interfaces' 
states:
\begin{Pro}
For an interface:
$$
\forall t\geq 0\ \left\{\begin{array}{ccccl}
|x_s^t|&\leq &a^tM&{\rm if}&s\leq J(x^0)-t-1\\
|1-x_s^t|&\leq &a^tM&{\rm if}&s\geq J(x^0)+t.
\end{array}\right.
$$
\end{Pro}
This result simply follows from the linearity of the CML and the inequality
\begin{equation}\label{INCRE}
|J(x^t)-J(x^0)|\leq t,
\end{equation}
which is a consequence of the nearest neighbors' coupling.
\bigskip

Combining the previous results, one can give some bounds for the temporal 
code of an interface
\begin{Pro}\label{SIGMA}
If $c\in {\rm I}_{p\over q}^0$ and $\{x^t\}_{t\geq 0}$ is an interface, then
$$\exists \sigma_1,\sigma_2\in\R\ :\ \forall t\geq 0\ 
\lceil{p\over q}t+\sigma_1\rceil\leq J(x^t)\leq 
\lceil{p\over q}t+\sigma_2\rceil .$$
\end{Pro}

\noindent
{\sl Proof:} We here prove the left inequality, the right one follows 
similarly. If ${\displaystyle{p\over q}}=-1$, 
then the statement holds by the inequality (\ref{INCRE}). Now, let $t_0$ be 
given by Remark \ref{T0}. For ${\displaystyle{p\over q}}\neq -1$ we have 
$$\left( R_0\right)_{-2t_0-1}>0.$$
Let $t_1$ be such that $a^{t_1}M\leq \left( R_0\right)_{-2t_0-1}$ and let
$\tilde\sigma_1=J(x^0)-t_1-2t_0-1$. Define
$$y_s=\left\{\begin{array}{ccl}
x_s^{t_1}&{\rm if}&s<\tilde\sigma_1-2t_0-1\\
\left( R_{\tilde\sigma_1}\right)_s&{\rm if}&\tilde\sigma_1-2t_0-1\leq s\leq 
\tilde\sigma_1+2t_0\\
max\left\{x_{J(x^{t_1})}^{t_1},x_s^{t_1}\right\}&{\rm if}& 
s>\tilde\sigma_1+2t_0.
\end{array}\right.$$
Let us check that $x^{t_1}\prec y$. For $\tilde\sigma_1-2t_0-1\leq s\leq 
\tilde\sigma_1+2t_0$, $y_s\geq \left( R_0\right)_{-2t_0-1}\geq x_s^{t_1}$ 
using the monotony of $R_{\tilde\sigma_1}$ and the previous proposition. 
\bigskip

Hence, by Proposition \ref{ITEOR}, we obtain $J(x^{t+t_1})\geq J(y^t)
\quad \forall 
t\geq 0$. But $y$ is a kink that satisfies the condition (\ref{CONDIN}) in
the framework of Remark \ref{T0}, 
therefore the left inequality is stated.
$\Box $
\bigskip

Finally, the statement on the velocity of any interface, valid for all
the values of the parameters, is given by
\begin{Thm}\label{INTVE}
If $\{x^t\}_{t\in\N}$ is an interface then, its velocity exists and
$$\lim_{t\rightarrow +\infty} {J(x^t)\over t}=\bar{v}(a,\epsilon,c).$$
\end{Thm}

\noindent
{\sl Proof:}  For any $\bar{v}$ given by (\ref{VELOCITY}), let 
${\displaystyle{p_1\over q_1}}\leq \bar{v},
\ {\displaystyle {p_2\over q_2}}\geq \bar{v},\ 
c_1\in {\rm I}_{p_1\over q_1}^0$ and 
$c_2\in {\rm I}_{p_2\over q_2}^0$ respectively. 
If $c\in {\rm I}_{p\over q}^0$, then the statement clearly 
follows from the previous proposition.
Moreover, by the monotony of the local map, we have
$$\forall x\in {\cal M}\quad
F_{\epsilon,c_2}x\prec F_{\epsilon,c}x\prec F_{\epsilon,c_1}x,$$
where the dependence on $c$ has been added to the CML. It clearly follows that
$$
{p_1\over q_1}\leq\liminf_{t\rightarrow +\infty}{J(F_{\epsilon,c}^tx)\over t}
\leq\limsup_{t\rightarrow +\infty}{J(F_{\epsilon,c}^tx)\over t}
\leq {p_2\over q_2}.
$$
The convergence is then ensured by choosing ${\displaystyle{p_1\over q_1}}$ and 
${\displaystyle{p_2\over q_2}}$ arbitrarily close to $\bar{v}$. $\Box $
\bigskip

To conclude this section, we mention that all these results and, in particular
Theorem \ref{INTVE}, can be extended to the orbits for which the initial 
condition satisfies
$$\lim_{s\rightarrow\pm\infty}H(x_s)-H(s)=0.$$
For such orbits, one can prove Theorem \ref{INTVE} for the following codes:
$$J_{inf}(x^t)=\min\{s: x_s^t\geq c\}\quad {\rm and}\quad
J_{sup}(x^t)=\max\{s: x_{s-1}^t< c\}.$$
Actually, we have 
$$\lim_{t\rightarrow +\infty}{J_{inf}(x^t)\over t}=
\lim_{t\rightarrow +\infty}{J_{sup}(x^t)\over t}=\bar{v}.$$

\section{The nonlinear stability of fronts with rational velocity}
An extension of the linear stability result for the 
fronts can be achieved, 
also using the CML contracting property which appears in (\ref{LINDY}),
by proving that for all the kink initial conditions, the corresponding 
interface's code is
identical to a front's code, at least after a transient. 

\subsection{The co-existence of fronts and non-uniform fronts}
We shall see below that one can only claim the nonlinear 
stability of fronts in some regions of the 
parameters. We now justify such restrictions on the latter. 
\bigskip

Let, a 
{\bf non-uniform front}, be an interface in $\Lambda$ for which the temporal
code cannot be written with the relation (\ref{TEMPCO}). Using the admissibility condition, one can show the existence of such orbits given a
non-uniform code. In this way, we prove in Appendix \ref{NONUNI}, the 
existence
of the non-uniform fronts with velocity ${1\over 2}$ and 0 respectively,
when $\epsilon$ is close to 1.
Since these orbits are interfaces, they co-exist with the fronts 
of the same velocity by Theorem \ref{INTVE}. Moreover, it is possible
to state the linear stability of these orbits similarly to Proposition \ref{STAB2}. Hence, when the non-uniform fronts exist, the fronts do not 
attract all the kink initial conditions.

\subsection{The nonlinear stability for weak coupling}
Here, we state the nonlinear stability's result for the fronts with
non negative rational velocity, in the 
neighborhood of $\epsilon=0$. However, the following assertion can similarly be
extended to the fronts with negative velocity.
\bigskip

Let 
$$\Delta_{p\over q}\equiv\eta({p\over q})-\eta({p\over q}^-)=
\sum_{k=1}^{+\infty}l_{kp.kq}$$
and, for $\theta\in (0,1)$ let the interval
$${\rm I}_{p\over q}^{\theta}=\left(\eta({p\over q}^-),
\eta({p\over q})-\theta\Delta_{p\over q}\right),$$
we have:
\begin{Thm}\label{GLOBST}
Given ${p\over q}\in \Q\cap [0,1]$, $a\in (0,1)$ and $\theta\in (0,1)$, there
exists $\epsilon_0>0$ such that for any kink initial condition $x$:
$$\forall \epsilon\in (0,\epsilon_0)\quad \forall c\in 
{\rm I}_{p\over q}^{\theta}\quad \exists\sigma\in\R\ :\ 
\lim_{t\rightarrow +\infty}\|R_{\sigma}^t-x^t\|=0.$$
\end{Thm}

\begin{Rem}\label{VIT0}
For the velocity ${p\over q}=0$, the statement still hold for $\theta=0$.
\end{Rem}
Moreover for the velocity ${p\over q}=1$, using similar techniques , we have 
proved that the theorem holds with $\theta=0$ and $\epsilon_0=1$ for all the 
interfaces.
 
\subsection{Proof of the nonlinear stability}
In the proof of Theorem \ref{GLOBST} we assume $\epsilon<{2\over 3}$ then, by 
Proposition 
\ref{CROIS}, all the orbits $\{x^t\}_{t\in\N}$ under consideration are 
interfaces.
\bigskip

In a first place, it is shown that the interfaces propagate forward provided 
some conditions on the parameters are satisfied.
\begin{Pro}\label{PROP}
It exists $\epsilon_1>0$ such that for any 
interface $\{x^t\}_{t\in\N}$ and any $c\geq {1\over 2}$, we have
$$\forall \epsilon\in (0,\epsilon_1)\quad \exists t_0\ :\ \forall 
t\geq t_0\quad J(x^{t+1})\geq J(x^t).$$ 
\end{Pro}

\noindent
{\sl Proof:} According to the present phase space, all the orbits of the CML 
are bounded in the following way:
$$\forall \delta>0\quad \exists t_0\ :\ \forall t\geq t_0\quad 
-\delta <x_s^t<1+\delta\quad \forall s\in\Z.$$
In particular, for an interface:
$$x_{J(x^t)-1}^{t+1}<a(1-{\epsilon\over 2})c+{\epsilon\over 2}+
{a\epsilon\delta\over 2}\quad \forall t\geq t_0.$$
Hence, if $\epsilon$ is sufficiently small, the statement holds for all
$c\geq {1\over 2}$. $\Box $
\bigskip

Furthermore, we introduce a convenient configuration. 
Then, after computing the code of the corresponding orbit, we show 
a dynamical relation between the code of any interface and the code of this 
particular orbit. 
\bigskip
 
Let
$$\forall j\in\Z\quad\left( S_j\right)_s=
{\displaystyle \sum_{k=1}^{+\infty} \sum_{n=-k}^k} 
l_{n,k} H(s-n-j)\quad \forall s \in \Z.$$
 Notice that if $c\in (\eta(0^-),\eta(0)]$, $S_j$ is the front of 
velocity 0. When $c\not\in (\eta(0^-),\eta(0)]$, this configuration
is a convenient tool. Actually, the temporal code for the orbit 
$\{S_0^t\}_{t\in\N}$ is shown to be given by:
\begin{Pro}\label{THETA}
Given ${p\over q}>0$, $a\in (0,1)$ and $\theta\in [0,1)$, there exists 
$\epsilon_0 >0$ such that
$$\forall \epsilon\in (0,\epsilon_0)\quad \forall c\in 
{\rm I}_{p\over q}^{\theta}\quad J(S_0^t)=\lceil (t+1){p\over q}\rceil
\quad\forall t\geq 0.$$
Moreover
$$\inf_{s,t}|(S_0^t)_s-c|>0.$$
\end{Pro} 
The proof is given in Appendix \ref{P-THETA}.
\bigskip

We now state the main property of an interface's code.
\begin{Pro}\label{PR85}
For an interface we have,
under the same conditions of the previous proposition
\begin{equation}\label{BORNE}
\exists n_0\quad \forall n\geq n_0\quad 
J(x^{t+n})\leq J(S_0^{t-1})+J(x^n)\quad\forall t\geq 1.
\end{equation}
\end{Pro} 

\noindent
{\sl Proof:} Using the relation (\ref{LINDY}) for $t=n-1$, we obtain the 
relation between the interfaces' states and the configuration $S_j$:
$$x^n=S_{J(x^{n-1})}+{\rm L}\left(x^{n-1}-S_{J(x^{n-1})}\right).$$
By induction, the latter leads to 
$$x^n=S_{J(x^{n-1})}
+{\displaystyle \sum_{k=m+1}^{n-1}} {\rm L}^{n-k}\left(S_{J(x^{k-1})}-
S_{J(x^k)}\right)+{\rm L}^{n-m}\left(x^m-S_{J(x^m)}\right)\quad
\forall n>m.$$
Then the monotony of $S_j$, the positivity of ${\rm L}$ and 
Proposition \ref{PROP} induce the following inequality for $m$ large enough and
 $n>m$
$$\tilde S_n\prec x^n,$$
where 
$$\tilde S_n=S_{J(x^{n-1})}+{\rm L}^{n-m}\left(x^m-S_{J(x^m)}\right).$$
Consequently
$$J(x^{t+n})\leq J(\tilde S_n^t)\quad \forall t\geq 0.$$
\bigskip

Now, given $\delta={\displaystyle \inf_{s,t}|(S_0^t)_s-c|}$
according to the previous Proposition, let $n_0>m$ be such that
$$a^{n_0-m}\|x^m-S_{J(x^m)}\|<\delta.$$
It follows that 
$$\|\tilde S_n-S_{J(x^{n-1})}\|<\delta \quad \forall n\geq n_0+1,$$
and hence, by Proposition \ref{STAB1}
$$\begin{array}{rl}
J(\tilde S_n^t)=&J(S_{J(x^{n-1})}^t)\\
=&J(S_0^t)+J(x^{n-1})\quad \forall t\geq 0,n\geq n_0+1.
\end{array}$$
 $\Box$
\bigskip

Finally, we state the result which ensure a sequence to be uniform.
\begin{Lem}\label{ENS}
Let $\{ j_n\}_{n\in\N }$ be an integer sequence which satisfies the 
conditions $\forall n\geq n_0$:

\noindent
(i) $j_{n+k}\leq\lceil {p\over q}k\rceil+j_n \quad \quad 
\forall k\geq 0,$

\noindent
(ii) $j_n\geq \lceil {p\over q}n+\sigma\rceil\quad $ for a fixed 
$\sigma\in\R$,

\noindent
then 
$$\exists \gamma\in\R\ {\rm and}\ n_1\geq n_0 : \forall n\geq n_1\quad 
j_n=\lceil {p\over q}n + \gamma\rceil .$$
\end{Lem} 
The proof is given in Appendix \ref{P-ENS}.
\bigskip

Collecting the previous results, we can conclude that,
under the conditions of Proposition \ref{PROP} and 
\ref{THETA}, the temporal code of any interface satisfies the condition
{\it (i)} of Lemma \ref{ENS} (Notice that Proposition \ref{PROP} only serves in
 the proof of Proposition \ref{PR85}). Moreover, this code also 
satisfies
the condition {\it (ii)} by Proposition \ref{SIGMA}. Hence, after a transient,
all the interfaces have a front's code. By (\ref{LINDY}), they consequently
converge to a front.
\bigskip

To conclude the proof of Theorem \ref{GLOBST} in the framework of Remark 
\ref{VIT0}, we let $c\in{\rm I}_0^0$. Hence $J(S_0^t)=0\quad \forall t$. 
If $c\geq {1\over 2}$, the statement holds using Proposition \ref{PROP} and the
relation (\ref{BORNE}). For $c<{1\over 2}$, one can prove the following 
inequalities:
$$\exists t_0\ : \ \forall t\geq t_0\quad J(x^{t+1})\leq J(x^t),$$
similarly to Proposition \ref{PROP} and, 
$$\exists n_1\ : \ \forall n\geq n_1\quad J(x^{n+1})\geq J(S_0)+J(x^n),$$
similarly to the relation (\ref{BORNE}).
 
\section{Concluding Remarks}
In this article, we have considered a simple space-time discrete model for 
the dynamics of various nonlinear extended systems out of equilibrium. The 
(piece-wise) linearity of this bistable CML allowed the construction of a 
bijection 
between the set of global orbits and the set of admissible 
codes. When they do not reach the discontinuity, these orbits are linearly 
stable. For 
$\epsilon < {1\over 2}$, the CML is injective on ${\cal M}$, then $\Lambda$ can
be identified with the limit set 
$$\bigcap_{t=0}^{+\infty}F_{\epsilon}^t({\cal M}),$$
which attracts all the orbits. These comments justify the study of the global 
orbits and in particular, the study of fronts which occur widely in extended 
systems.
\bigskip

The existence of fronts, with a parametric dependence for their velocity, has 
been proved using the spatio-temporal coding. The velocity was shown to be 
increasing with $c$. We have in addition checked numerically that $\bar{v}$ is 
also an increasing function of 
$\epsilon$ on $(0,{1\over 2})$ for any $a$ and on $(0,1)$ for any $a\in 
(0,{1\over 2})$ or for any $a$ if $c$ is such that $\bar{v}(a,1,c)=1$.
Moreover, one can find some values of $a$ and  $c$ for which the 
front velocity does not (always) increase with $\epsilon$ (see Figure 2). The 
spatio-temporal coding also serves to show the existence of other patterns 
such as non-uniformly propagating interfaces (Appendix \ref{NONUNI}). When 
these exist, they always co-exist with the fronts of the same velocity.
\bigskip

Furthermore, we have consider the more general dynamics of interfaces.
Using the temporal code, we have shown that all these orbits have the front's 
velocity uniquely determined by the CML parameters.
\bigskip

The stability of fronts was also proved, firstly with respect to 
initial conditions close to a front state in their "center", and secondly 
with respect to any kinks for the fronts with non negative rational velocity,
assuming some restrictions on the parameters. Actually, the latter allows us to 
avoid the existence of non-uniform fronts which would attract certain 
interfaces.
\bigskip

Finally, notice that all the results stated for the orbits that never 
reach the discontinuity can be extended to some CML's with a $C^{\infty}$ 
bistable local map.
Actually, one can modify the local map into a $C^{\infty}$ one, in the 
intervals where the orbits in $\Lambda$ never go, without changing these 
orbits. In other terms, all our results stated in open sets can 
be extended to differentiable maps, in particular, the existence and the 
linear stability of fronts and non-uniform fronts with rational velocity.
\bigskip

These last results show the robustness of fronts in these models. This
emphasizes the realistic nature of these models.
\bigskip

\noindent
{\bf Acknowledgments}

\noindent
We want to thank R.\ Lima, E.\ Ugalde, S.\ Vaienti and R.\ Vilela-Mendes for 
fruitful discussions and relevant comments. We also acknowledge L.\ Raymond
for his contribution to the nonlinear stability's proof and, P.\ Collet for
attracting our interest to this piece-wise affine model.

\vfill\eject

\appendix
\section{Properties of the coefficients $l_{n,k}$}\label{LNK}

\noindent
In first place, if $k\geq 1$ the following expression holds 
$$l_{n,k}=(1-a)a^{k-1}(-1)^n\sum_{p=|n|}^k\left({-\epsilon\over 2}\right)^p
{k\choose p}{2p\choose p+n}$$
where ${n\choose p}$ stands for the binomial coefficients. In particular, one
has 
$$\lim_{\epsilon\rightarrow 0}l_{n,k}=(1-a)a^{k-1}\delta_{n,0},$$
and 
$$l_{n,k}=\left\{\begin{array}{ccc}
{1-a\over a}({a\over 2})^k{k\choose {k+n\over 2}}&{\rm if}
&{k+n\over 2}\in \Z\\
0&{\rm if}&{k+n\over 2}\not\in \Z
\end{array}\right. \quad {\rm if}\ \epsilon=1.$$
\bigskip

Moreover, one can show that these coefficients have the following generating 
function
$${1-a\over a}\left({a\epsilon\over 2}+a(1-\epsilon)x+{a\epsilon\over 2}x^2
\right)^k=\sum_{n=0}^{2k}l_{n-k,k}x^n\quad \forall x\in \R,\ k\geq 1.$$
From this function we deduce the relation
$$
\sum_{m\in\Z }l_{n-m,t}l_{m,k}={1-a\over a}l_{n,t+k}\quad 
\forall t,k\geq 1. \eqno(A1)
$$
Finally, we prove the following behaviour when $\epsilon$ is small
\begin{Pro}\label{APPEN}
$\forall q\geq 1$ and $q\geq p\geq0$

\noindent
(i) ${\displaystyle \lim_{\epsilon\rightarrow 0}}{l_{p,q}\over \epsilon^p}>0.$

\noindent
(ii) ${\displaystyle \lim_{\epsilon\rightarrow 0} 
{1\over \epsilon^p}\sum_{k=q+1}^{+\infty}
\sum_{n=p+1}^k} l_{n,k}=0.$
\end{Pro}

\noindent
{\sl Proof:} Assume $q\geq 1$ and $p\geq0$. {\it (i)}, From the explicit 
expression of the $l_{n,k}$, we have
$$\lim_{\epsilon\rightarrow 0} {l_{p,q}\over \epsilon^p}=
{(1-a)a^{q-1}\over 2^p} {q\choose p}.$$ 
Moreover, {\it (ii)} is ensured by the relation:
$$\sum_{k=1}^{+\infty}\sum_{n=p+1}^k l_{n,k} =\epsilon^{p+1} 
{a^p\over (1-a)^{p+1}}
{2^p\over \sqrt{1+{2a\epsilon\over 1-a}}}
\left({1\over 1+\sqrt{1+{2a\epsilon\over 1-a}}}\right)^{2p+1}.$$
$\Box$

\section{The existence of non-uniform fronts}\label{NONUNI}
\subsection{The non-uniform front with velocity ${1\over 2}$}
Consider the following sequence:
$$j_t=\lfloor {t\over 4}\rfloor+\lfloor {t+1\over 4}\rfloor\quad 
\forall t\in\Z,$$
and 
$$z_s^t=\sum_{k=1}^{+\infty}\sum_{n=-k}^k l_{n,k} H(s-n-j_{t-k})\quad 
\forall s,t\in \Z $$
the corresponding orbit in $\Lambda$, namely a non-uniform front 
with velocity ${1\over 2}$. Actually, $j_{t+4}=j_t+2$ but 
it does not correspond to the temporal code of the front of velocity 
${1\over 2}$. The existence of this particular orbit is stated in
\begin{Pro}
$\exists \tilde\epsilon >0$ such that $\forall\epsilon\in (\tilde\epsilon ,1]$,
there exists an interval $\tilde {\rm I}$ so that
$$\forall c\in \tilde {\rm I}\quad \forall t\in\Z\quad F_{\epsilon}(z^t)=
z^{t+1}.$$ 
\end{Pro}

\noindent
{\sl Proof:} $z_s^t$ is monotone in $s$. Then one has to prove that, in a 
neighborhood of $\epsilon=1$, 
$$\min_t z_{j_t}^t>\max_t z_{j_t-1}^t,$$
in order for the statement to hold for $\tilde {\rm I}=
\left({\displaystyle \max_t} z_{j_t-1}^t,
{\displaystyle \min_t} z_{j_t}^t\right]$.
\bigskip

Let $\Delta_k^t=j_t-j_{t-k}$. We have
$$\Delta_k^{t+4}=\Delta_k^t\quad {\rm and}\quad \Delta_{k+4}^t=\Delta_k^t+2,$$
and
$$\min_t \Delta_k^t=\Delta_k^2\quad {\rm and}\quad 
\max_t \Delta_k^t=\Delta_k^0.$$
Using these properties, we show that 
$$
\min_t z_{j_t}^t-\max_t z_{j_t-1}^t=\sum_{k=1}^{+\infty}(-1)^k l_{k,2k}.
$$
From the expression of the $l_{k,2k}$ for $\epsilon=1$, it follows that
$$\lim_{\epsilon\rightarrow 1^-} (\min_t z_{j_t}^t-\max_t z_{j_t-1}^t)>0.$$
$\Box $

\noindent
Notice that (when it exists) this non-uniform front lies between the fronts 
of code $\lfloor {t-1\over 2}\rfloor$and $\lfloor {t\over 2}\rfloor$.

\subsection{The oscillating front}
Here, we consider a non-uniform front of velocity 0. In particular, the one
for which the temporal code writes:
$$J_t={1+(-1)^t\over 2}\quad \forall t\in \Z.$$
Naturally, the expression for this orbit can be written similarly to the previous one but now, with this code's relation. Let $\{Z^t\}_{t\in\Z}$ denotes
the orbit. Its existence is claimed in
\begin{Pro}
$\exists \check\epsilon >0$ such that $\forall\epsilon\in (\check\epsilon ,1]$, 
there exists an interval $\check {\rm I}$ so that
$$\forall c\in \check {\rm I}\quad \forall t\in\Z\quad F_{\epsilon}(Z^t)=
Z^{t+1}.$$ 
\end{Pro}
Similarly to the previous proof, the existence of this orbit is ensured 
by the relation
$$
\min_t Z_{J_t}^t-\max_t Z_{J_t-1}^t=\sum_{k=1}^{+\infty}(-1)^k l_{0,k}.
$$

\bigskip 

Furthermore, using these technique, we can prove the existence of another 
non-uniform fronts. All these orbits exist only in a neighborhood of $\epsilon=
1$, they disappear for $\epsilon$ close to 0.

\section{Proof of Proposition 8.4}\label{P-THETA}
Since the fronts' configurations are
increasing functions in space, the following ordering holds:
$$S_0\prec R_{p\over q}.$$
Hence, using Proposition \ref{ITEOR}, we obtain for 
$c\in {\rm I}_{p\over q}^{\theta}$:
$$\left(S_0^t\right)_{\lceil {p\over q}(t+1)\rceil -1}\leq 
\left(R_{{p\over q}(t+1)}\right)_{\lceil {p\over q}(t+1)\rceil -1}<c
\quad \forall t\geq 0.$$
To prove the statement, we then show by induction that
$$\left(S_0^t\right)_{\lceil {p\over q}(t+1)\rceil }> c\quad 
\forall t\geq 0 ,$$
in a neighborhood of $\epsilon=0$.
\bigskip

Using the relations (\ref{CONFIG}) and (A1) to write 
$$\left(S_0^t\right)_s=\sum_{k=1}^t\sum_{n=-k}^k l_{n,k}H(s-n-J(S_0^{t-k}))
+\sum_{k=t+1}^{+\infty}\sum_{n=-k}^k l_{n,k}H(s-n),$$
we consider the following decomposition
$$\left(S_0^t\right)_{\lceil {p\over q}(t+1)\rceil }
-\eta({p\over q})+\theta\Delta_{p\over q}=P_1+P_2+P_3+P_4,$$
where
$$
\begin{array}{rcl}
P_1&=&{\displaystyle\sum_{k=1}^t\sum_{n=-k}^k}l_{n,k}
\left(H(\lceil{p\over q}(t+1)\rceil-\lceil{p\over q}(t-k+1)\rceil-n)
-H({p\over q}k-n)\right)\\
P_2&=&{\displaystyle\sum_{n=-t-1}^{t+1}}l_{n,t+1}
\left(H(\lceil{p\over q}(t+1)\rceil-n)-H({p\over q}(t+1)-n)\right)\\
P_3&=&{\displaystyle\sum_{k=t+2}^{+\infty}\sum_{n=-k}^k}l_{n,k}
\left(H(\lceil{p\over q}(t+1)\rceil-n)-H({p\over q}k-n)\right)\\
P_4&=&\theta{\displaystyle\sum_{k=1}^{+\infty}}l_{kp,kq}.
\end{array}
$$
We have the following inequalities:

\noindent
$\begin{array}{ccl}
P_1&\geq&0\quad {\rm since}\quad 
\lceil{p\over q}(t+1)\rceil-\lceil{p\over q}(t-k+1)\rceil
\geq \lfloor {p\over q}k\rfloor\ \forall k,t\in\N .\\
P_2&\geq&0\quad {\rm and}\quad 
P_2=l_{\lceil {p\over q}(t+1)\rceil,t+1}\quad {\rm if}\quad t<q-1.\\
P_3&\geq&{\displaystyle\sum_{k=t+2}^{+\infty}\sum_{n=-k}^k}l_{n,k}
\left(H(\lceil{p\over q}(t+1)\rceil-n)-H(k-n)\right)\\
 &\geq&-{\displaystyle\sum_{k=t+2}^{+\infty}
\sum_{n=\lceil{p\over q}(t+1)\rceil+1}^k}l_{n,k}.\\
P_4&\geq& \theta l_{p,q}.
\end{array}
$

\noindent
Then one can write, for $0\leq t<q-1$:
$$P_1+P_2+P_3+P_4\geq l_{\lceil {p\over q}(t+1)\rceil,t+1}-
\sum_{k=t+2}^{+\infty}\sum_{n=\lceil{p\over q}(t+1)\rceil+1}^kl_{n,k},$$
and for $t\geq q-1$:
$$P_1+P_2+P_3+P_4\geq\theta l_{p,q}-\sum_{k=q+1}^{+\infty}\sum_{n=p+1}^k 
l_{n,k}.
$$
Hence using Proposition \ref{APPEN}, in both cases $P_1+P_2+P_3+P_4>0$ 
if $\epsilon$ is sufficiently small. $\Box $

\section{Proof of Lemma 8.6}\label{P-ENS} 
By {\sl (ii)}, it exists $n_1\geq n_0$ such that 
$$j_{n_1}-n_1{p\over q}\leq j_n-n{p\over q}\quad \forall n\geq n_1.$$
Then, also using {\sl (i)}, one has
$$0\leq j_{n_1+t}-j_{n_1}-t{p\over q}\leq \lceil t{p\over q}\rceil 
-t{p\over q}<1\quad \forall t\geq 0,$$
which implies that
$$\lfloor j_{n_1+t}-j_{n_1}-t{p\over q}\rfloor =0\quad \forall t\geq 0.$$
Now it is straightforward to see that
$$j_n=\lceil n{p\over q}+\gamma\rceil\quad \forall n\geq n_1,$$
where $\gamma=j_{n_1}-n_1{p\over q}.\Box$

\vfill\eject

\bibliographystyle{unsrt}





\vfill\eject

\begin{figure}[h]
\epsfxsize=15cm
\centerline{\epsfbox{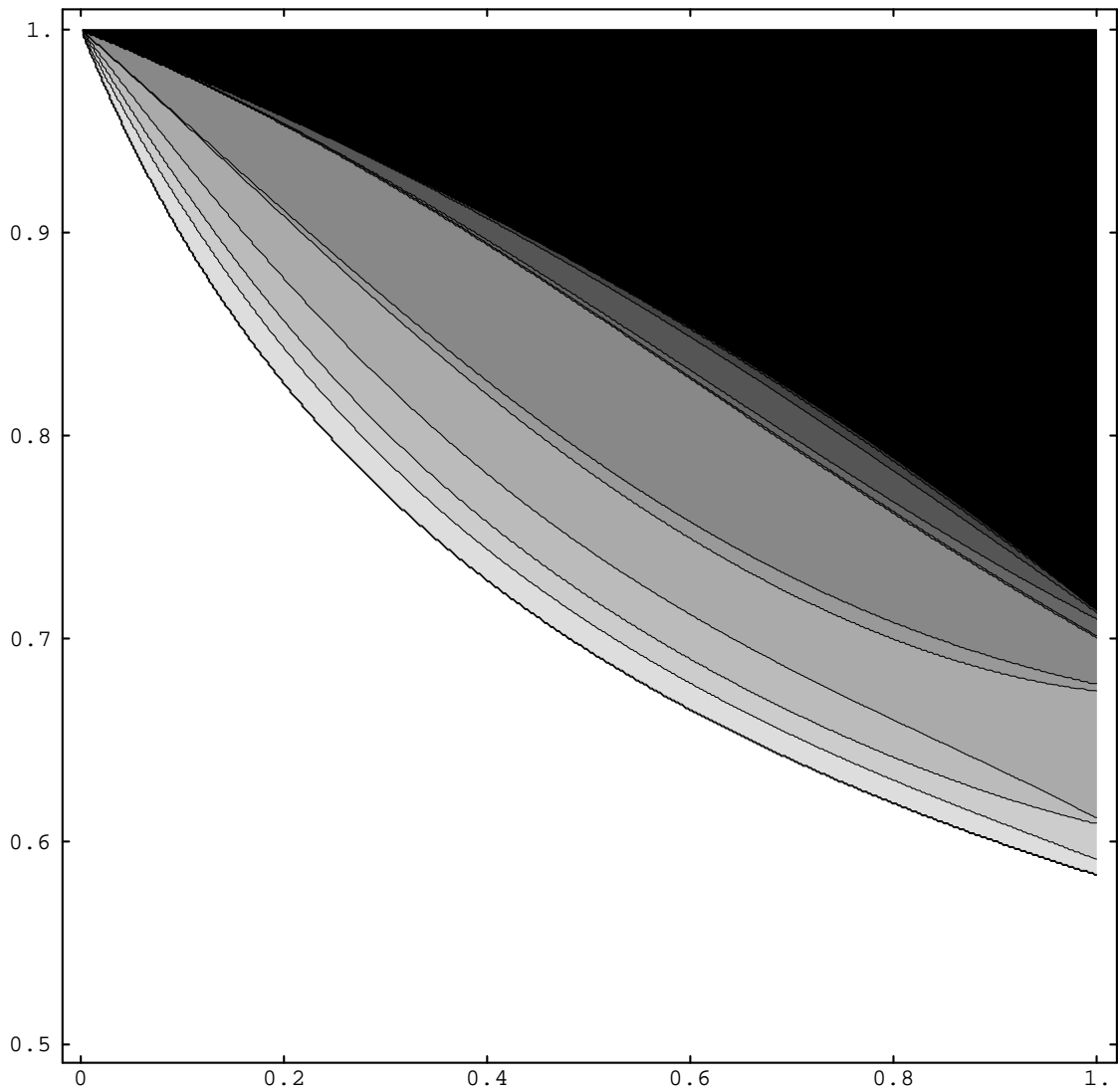}}
\caption{Grey-level plot of the front velocity versus 
$\epsilon$ (Horizontal axis) and $c$ (Vertical axis) for $a=0.6$ in range 
0 (white) to 1 (black). This figure is a projection of a two dimensional 
Devil's staircase.}
\end{figure}

\vfill\eject

\begin{figure}[h]
\epsfxsize=15cm
\centerline{\epsfbox{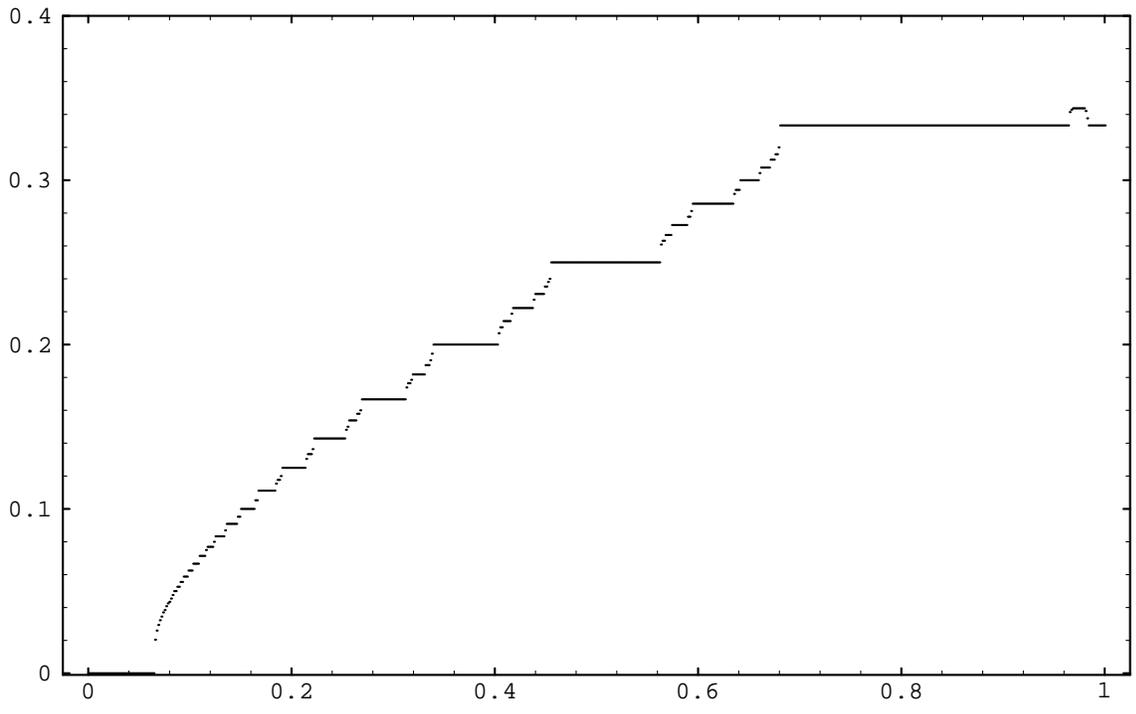}}
\caption{Plot of the front velocity versus $\epsilon$ for 
$a=0.9$ and $c=0.8216$.}
\end{figure}



\end{document}